\documentstyle[twocolumn,aps]{revtex}
\newcommand{\be}{\begin{equation}}
\newcommand{\ee}{\end{equation}}
\newcommand{\bea}{\begin{eqnarray}}
\newcommand{\eea}{\end{eqnarray}}

\newcommand{\sh}{Swift-Hohenberg equation}
 
\input epsf

\begin{document}
\pagestyle{plain}
\bibliographystyle{/usr/local/lib/tex/macros/prsty}
\title{Oscillon-type Structures and Their Interaction in a Swift-Hohenberg Model}
\author{Catherine Crawford\thanks{Corresponding author.  Tel: (847) 491-3345;
Fax: (847) 491-2178; E-mail:  ccrawford@nwu.edu.} and Hermann Riecke}
\address{Department of Engineering Sciences and Applied Mathematics,
Northwestern University, Evanston, IL 60208, USA}
\date{30 April 1998}
\maketitle 

\begin{abstract}

Motivated by the observation of localized circular excitations (`oscillons')
in vertically vibrated granular layers\cite{UmMe96}, we numerically investigate
an extension of a Swift-Hohenberg model that exhibits a subcritical transition
to square patterns. For sufficient subcriticality, stable
oscillon structures
are found.  The localization mechanism is quite general and is due to
non-adiabatic effects.   
Bound structures of oscillons of equal and opposite
polarity are found with bound states of like
polarity being less robust.
Much of the phenomena are consistent with the experimental observations and
suggest that oscillons are not specific
to patterns in granular media or to parametrically driven systems.  
Experimental tests are suggested that would determine whether this minimal framework is
sufficient to describe the phenomena.\\

\end{abstract}

\section{Introduction}

Recent experiments on vertically vibrated thin layers of granular material have revealed a wealth of
interesting phenomena \cite{UmMe96,MeUm94,MeUm95,UmMe97,JaNa96,MeKn97}. Depending on the driving frequency and amplitude
different types of standing-wave patterns are observed. They can be ordered, e.g.
stripe patterns and square patterns as they are also observed in parametrically excited 
surface waves on liquids, or disordered. Particularly exciting is the observation of 
two-dimensional localized structures which were termed `oscillons' \cite{UmMe96}. 
Due to the subharmonic response of the excitation,
they resemble during one cycle of the driving a circular peak and during the next
cycle a circular crater. Since the driving is invariant under shifts by one period,
 peaks and craters
can exist simultaneously in any given cycle. The excitations are found to form bound structures.
Thus in the experiments, stable pairs, tetramers, and chains of alternating
polarity are observed. In fact, even the square pattern 
can be thought of as being built up from these individual units rather than extended waves. 

Stimulated by these experiments oscillon-type patterns have been studied theoretically 
from a number of different angles \cite{TsAr97,CeMe97,Ro97,VeOt97}. 
In \cite{TsAr97} oscillons are found within a 
Ginzburg-Landau approach; assuming small amplitudes and weak damping
of the waves a Ginzburg-Landau equation for the wave amplitude is coupled to a mode describing the
local height of the granular layer. In \cite{VeOt97} the waves are described using a
stroboscopic map in which the waves arise in a period-doubling bifurcation. 
A more detailed modelling of the dynamics of the granular layer is pursued in 
\cite{CeMe97,Ro97,VeOt97,Sh97,EgRi98}.
In \cite{Ro97} the granular layer is modelled as a sheet that locally follows the dynamics
of an inelastic ball whereas in 
\cite{CeMe97} the core ingredients
are the free flight of the granular particles and diffusion. While in 
 \cite{TsAr97,CeMe97,Ro97,VeOt97} oscillon-type structures have been found numerically, 
the mechanism that is responsible
for the localization of the waves has not been clearly identified.

In this paper we wish to emphasize two main points.
First, we demonstrate that oscillon-type patterns are neither specific to parametrically
excited waves nor to granular media. By studying a suitable extension of the Swift-Hohenberg
equation we show that stable oscillons may be expected quite generally in systems that undergo
a sufficiently hysteretic transition to square patterns. The mechanism of localization is provided
by non-adiabatic effects \cite{Po86,BeSh88,Ri86,SaBr96}. In a very recent study of a somewhat
simpler Swift-Hohenberg equation \cite{SaBr97b} similar results were found.  
We present here a detailed investigation of the range of existence of oscillons and
their bound states.  Second, we compare our general results with the experimental observations
and comment on what may be expected in additional experiments.  This will help 
determine whether in this specific system additional mechanisms related to granular media are important,
such as a coupling to an additional field \cite{TsAr97}.

\section{The Model}

To study the connection between subcritical square patterns and oscillons we study a simple
Swift-Hohenberg-type model.
Central for the richness of states observed in connection with oscillons is the fact that 
there are two types of oscillons with equal and opposite polarity. The two types 
are related by a reflection symmetry that arises from the discrete time-translation symmetry
of the system. Within the framework of a Swift-Hohenberg equation this leads to the 
requirement that the equations be equivariant under the reflection of the amplitude. We therefore
consider the following two-dimensional equation for the amplitude $\psi$ that characterizes
the pattern,
\begin{eqnarray}
\partial_t \psi & = & R \psi - (\partial_x^2+1)^2 \psi +  b \psi^3 - c \psi^5 \nonumber \\
 & & \mbox{} + e \nabla\cdot[(\nabla\psi)^3]
 - \beta_1 \psi (\nabla \psi)^2 - \beta_2 \psi^2 \nabla^2 \psi.\label{e:sh}
\end{eqnarray}
As in the usual \sh \ the basic state $\psi=0$ becomes first unstable to a spatially periodic
structure with wave number $q=1$ at $R=0$. The choice of nonlinear terms is guided by the 
following considerations. Due to the reflection symmetry $\psi \rightarrow -\psi$ only 
odd powers of $\psi$ are possible. In order to obtain a subcritical bifurcation the usual 
\sh \  has to be
extended to fifth order. If only the nonlinearities $\psi^3$ and $\psi^5$ are present no 
stable square patterns arise. The term proportional to $e$ is chosen because it is 
known to favor square patterns \cite{GeSi81,CrHo93} and the terms proportional to the $\beta_i$'s allow
additional tuning possibilities.

Models of type (\ref{e:sh}) have been used to describe a wide range of patterns from
Rayleigh-B\'{e}nard convection to wide-area lasers.
The connection with the experiments on vibrated granular layers can be understood in the following way.
In a weakly nonlinear framework one attempts to describe the standing waves
by an expansion based on the mode that destabilizes the basic
state with undeformed surface, $h=0$. The surface deformation of a spatially
periodic standing wave
arising in a subharmonic instability can then be written as
\be
h({\bf r},t)=A\,H_\omega(t)\,e^{i\omega t/2} e^{i{\bf q}\cdot{\bf r}} +c.c. + h.o.t.,\label{e:expand}
\ee
where $H_\omega(t)$ has the same period as the driving, $2\pi/\omega$, and $A$ is the small
amplitude of the wave. In a Ginzburg-Landau approach the amplitude $A$ would be allowed to 
vary slowly in space and time. In order to capture the non-adiabatic effects that couple the 
slow and fast space dependence, no slow space dependence can be introduced. The order
parameter $\psi$ is therefore allowed to vary on a fast spatial scale, which incorporates the term 
$e^{i{\bf q}\cdot{\bf r}}$.  
The fast time dependence, however, is locked to the driving and can be 
kept outside the order parameter . From (\ref{e:expand}) it is
apparent that the symmetry $t \rightarrow t+2\pi/\omega$ induces the symmetry 
$\psi ({\bf r},t) \rightarrow -\psi ({\bf r},t)$ of the \sh .  Note that subharmonic instabilities
are characterized by a single real Floquet multiplier, $\mu = -1$.  Therefore, near onset the
order parameter $\psi$, describing the excited waves, is real.
An explicit derivation of a one-dimensional version of (\ref{e:sh}) from a periodically 
forced nonlinear Schr\"{o}dinger equation 
has been performed
in the context of phase-sensitive amplifiers of optical solitons \cite{KuHi94} 
\footnote{In \protect \cite{KuHi94} $e = 0$ and $(\partial_x^2+1)^2\psi$ is replaced by 
$(\partial_x^2-1)^2\psi$.}.
In that analysis damping is strong and the (strong) driving occurs 
periodically in a shock-like manner.
This situation is very similar to that encountered in the granular system.

To obtain an overview of the periodic solutions of (\ref{e:sh}) we perform a weakly nonlinear
analysis.  The ansatz $\psi = Ae^{i x} + Be^{i y} + c.c. + h.o.t.$ 
yields the following amplitude equation
\begin{eqnarray}
\partial_t A & = & R A - \beta \mid A\mid ^2 A - \gamma \mid B\mid^2 A \nonumber \\
& & \mbox{} - \delta \mid A\mid ^4 A 
   - 2 \rho \mid A\mid ^2\mid B\mid ^2 A - \rho \mid B\mid ^4 A, \label{e:amp}
\end{eqnarray}
where $\beta = -3b + 3e + \beta_1 -3 \beta_2, \gamma = -6b + 2e  + 2\beta_1 - 6\beta_2, 
\delta = (-3 b \beta_1 -9 e \beta_1 + 5 \beta_1^2 -11\beta_1\beta_2 + 480 c d)/48d$, 
and $\rho = (-3 b \beta_1 -e\beta_1 + 3 \beta_1^2 -7\beta_1\beta_2 +120 c d )/4d$.
The equation for $B$ is obtained by interchanging $A$ and $B$ in (\ref{e:amp}).
Thus the bifurcation to stripes and squares is subcritical for $\beta < 0$ and $ \beta + \gamma < 0$,
respectively.  Motivated by the experiments, we focus on the regime in which the bifurcation
to square patterns is subcritical.  Unless otherwise noted, the bifurcation to stripes is 
also subcritical.

\section{Oscillons}

\begin{figure}[ht]
\centerline{\epsfxsize=2.8in{\epsfbox{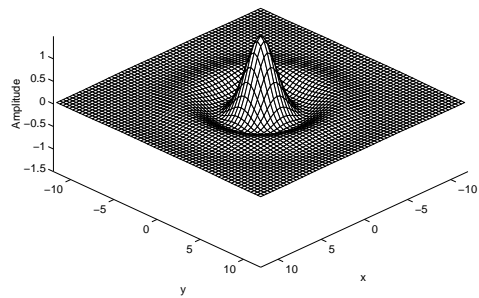}}}
\caption{ Single oscillon for the parameters are $R = -0.5, b = c = 1.0, e = 0.75, \beta_1 = \beta_2 = 2.0$.}   
\label{fig:oscillon}
\end{figure}

To investigate localized, oscillon-type solutions we study (\ref{e:sh}) numerically using
a pseudo-spectral code.  A single oscillon can be seen in Fig.~\ref{fig:oscillon}.
Fig.~\ref{fig:esaddle} shows the region found numerically for the existence of 
localized states for several values of $e$ for the case $\beta_1 = \beta_2$.
Below the solid triangles, oscillons do not exist, whereas
above the solid circles and squares the oscillons lose stability to 
extended stripes or squares, respectively.  Also shown are the saddle-node bifurcations for
squares and stripes.

\begin{figure}[ht]
\centerline{\epsfxsize=2.5in{\epsfbox{figures/esaddle/bbox/esn.bps}}}
\caption{ Numerically obtained saddle-node bifurcations for squares (open squares), stripes 
(open circles), and oscillons (solid triangle) along with the parameter values at
which oscillons lose stability to extended squares (solid squares) or stripes (solid circles).
The parameters are $b = c = 1.0, \beta_1 = \beta_2 = 2.0$.}
\label{fig:esaddle}
\end{figure}

We now discuss the mechanism involved in localizing these structures.  Consider an oscillon
as an extremely small domain of a patterned state connected by fronts to the basic, unpatterned
state.  Within the framework of a Ginzburg-Landau equation, fronts are stationary only for
a single parameter value.  In this setting, the interaction of fronts is attractive and such a 
localized state is unstable (e.g.,\cite{MaNe90,HaPo91}).  In \cite{Po86} 
the pinning of interfacial fronts by the interaction with the underlying 
small-scale periodic pattern has been pointed out.  The pinning effect is examined 
in a more quantitative analysis in \cite{BeSh88}.
It is found that the front is locked for a range of the control parameter which
is found to scale as $e^{-\alpha/ \sqrt{R}}$.
This same pinning was used to explain the existence of localized pulses in a 1-D
quintic Swift-Hohenberg equation \cite{SaBr96}.  
We expect the same mechanism to operate in two spatial dimensions.  
Hence, we also expect the range of existence of the oscillons to scale in a similar manner.

\begin{figure}[ht]
\centerline{\epsfxsize=2.5in{\epsfbox{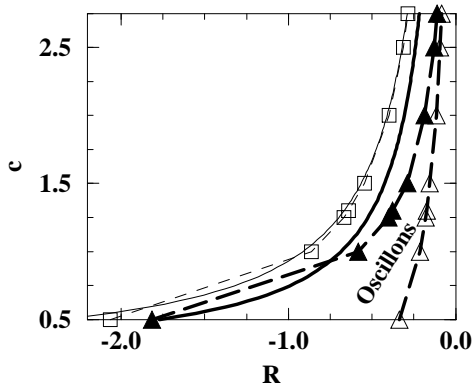}}}
\centerline{\epsfxsize=2.5in{\epsfbox{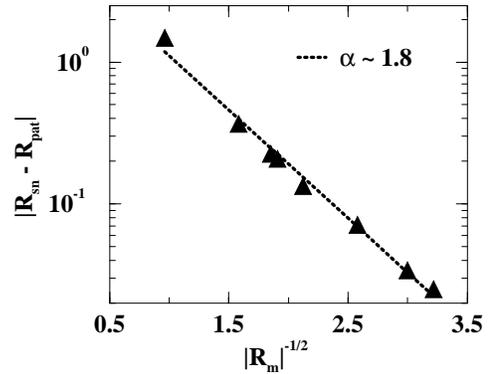}}}
\caption{(a) Oscillons exist between the solid triangles
and the open triangles.  The numerical saddle-node bifurcation for squares (open squares) along
with the analytical saddle-node curve  (thin solid line) are shown.  The thick solid line shows
the analytical curve where $\Delta E = 0$ (cf. \protect \ref{e:lyap})). (b) The existence range of a
single oscillon is plotted as a function of the inverse square root of the midpoint of the range.
The dashed line is a fit to $\kappa e^{-\alpha / \protect{\sqrt{R_m}}}$.
The parameters are $b=1.0, e = 0.75, \beta_1 = \beta_2 = 2.0$.}
\label{fig:range}
\end{figure}

To confirm that the pinning is responsible for the localization, we investigate the dependence
of the existence range of oscillons as a function of $c$.  Increasing the prefactor, $c$,
of the quintic term causes both squares and stripes to become less subcritical.
Fig.~\ref{fig:range}a shows a phase diagram for the
range of existence of the localized structures along with the numerical and analytical 
saddle-node curves for squares.  Since squares become less subcritical, an increase in 
$c$ necessitates an increase
in $R$ for finding oscillons.  The oscillons have oscillatory tails
that decay exponentially outward.  As $R$ is increased, the spatial decay of these tails
becomes more gradual producing larger radial oscillations.  Hence, the front locking 
becomes weaker and the localized solution loses stability to a periodic pattern.
To demonstrate the scaling of the locking range, we plot the range in the control
parameter $R$ where a single oscillon exists as a function of the inverse square root 
of the midpoint of this range.
The log-linear plot in Fig.~\ref{fig:range}b shows that the range does scale
proportional to $e^{-\alpha/ \sqrt{R}}$.

For $\beta_1 = \beta_2 = 2f$ equation (\ref{e:sh}) possesses a 
Lyapunov Functional ${\cal L}$,
\begin{eqnarray}
{\cal L} & = & \int dx \{-\frac{1}{2}R\psi^2 + \frac{1}{2}[(\nabla^2 + 1)\psi]^2 - \frac{1}{4} b \psi^4 + 
\frac{1}{6} c \psi^6 \nonumber \\
& & \mbox{} - \frac{1}{4} e (\nabla\psi)^4 + \frac{1}{4} f \psi^2 \nabla^2(\psi^2)\} \equiv \int dx E(x) \label{e:lyap}
\end{eqnarray}
with $\psi_t = -\delta {\cal L}/\delta \psi$.
In this case the `energy' difference $\Delta E$ between different states of the system
governs the motion of a single front.
For the stable zero state $E$ is 0, while for squares
it is $-RA^2/2 + (\beta + \gamma)A^4/4 + (\delta + 3\rho)A^6/6$ within the weakly nonlinear theory
(\ref{e:amp}).
The values of $R$ at which both states have the same energy ($\Delta E = 0)$ and at which a single
front would be stationary is shown as the thick curve in Fig.~\ref{fig:range}a.
The existence range of oscillons lies to the right of this curve, corresponding
to parameters for which the square pattern has a lower energy.  We attribute this difference 
to the attractive interaction between the fronts making up the oscillon. 

\begin{figure}[ht]
\centerline{\epsfxsize=1.5in{\epsfbox{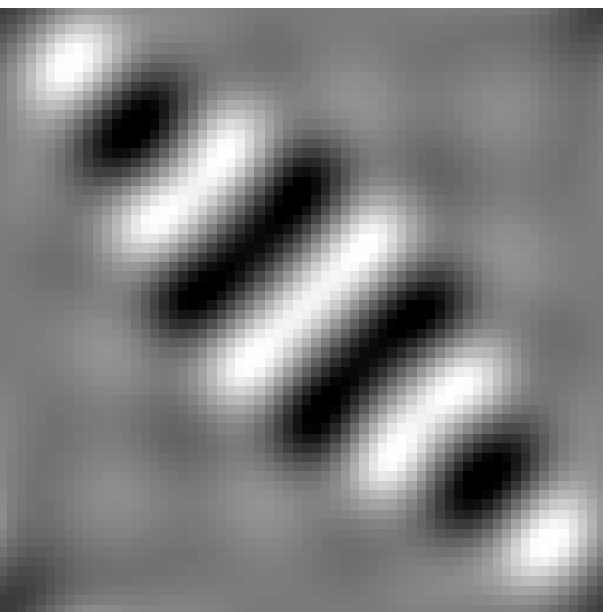}}\hspace{.25in}
       \epsfxsize=1.5in{\epsfbox{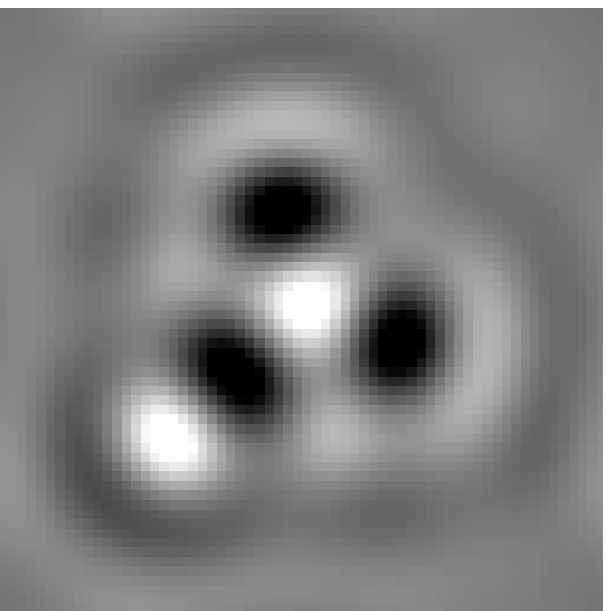}}}
\caption{The transition to stripes (a) and squares (b) starting from a single oscillon in the center.
         The parameters are $b = c = 1.0, \beta_1 = \beta_2 = 2.0$ and for (a) $R = -0.264, e = 0.5$
	 and (b) $R = -0.205, e = 0.75$.
\protect{\label{fig:trans}}
}
\end{figure}

When the oscillon loses stability to an extended pattern
state, stripes or squares emerge depending on other parameters.  Fig.~\ref{fig:trans} shows
intermediate stages for a single oscillon in the center evolving to stripes and squares, 
respectively.  As can be seen in the transition to stripes, the oscillon tends 
to elongate in one direction and generate small stripes in the other
direction, eventually filling the entire space with stripes.  
Note that the square pattern, although being unstable to stripes, still exists in this regime 
(cf. Fig.~\ref{fig:esaddle}).
The transition to squares is generally characterized by more oscillons appearing
on the perimeter of the initial oscillon.  These features seen in the numerical simulations 
are similar to the transitions observed in the experimental granular system.

\begin{figure}[ht]
\centerline{\epsfxsize=3.3in{\epsfbox{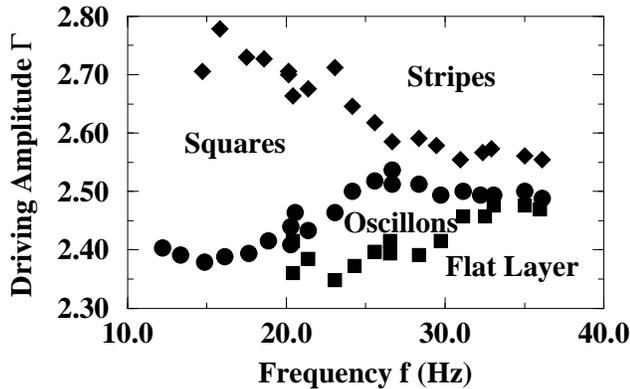}}}
\caption{The experimental phase diagram \protect{\cite{UmMe96}}.  For increasing 
$\Gamma$, patterns appear at the solid diamonds. As $\Gamma$ is decreased
the extended patterns exist until the solid circles beyond which oscillons exist in the mid-frequency range
until the solid squares.}
\label{fig:phasediag}
\end{figure}

\begin{figure}[ht]
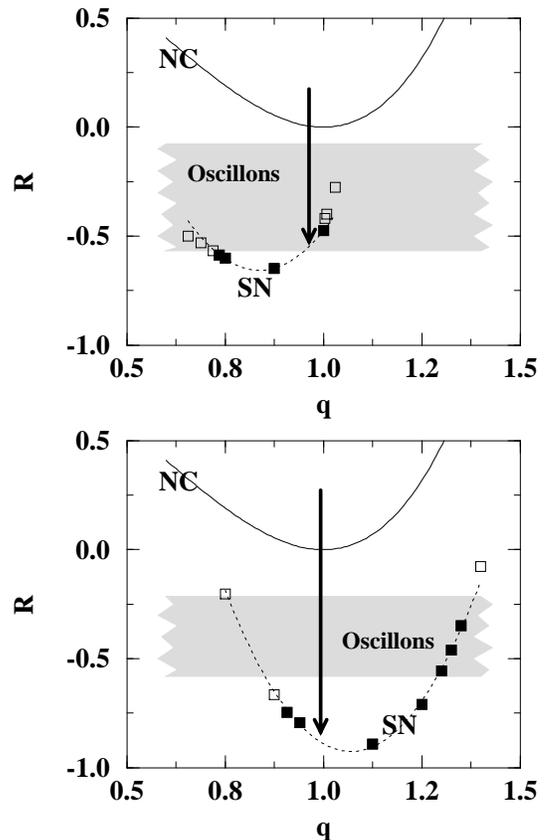

\centerline{\epsfxsize=2.8in{\epsfbox{figures/qcurves/bbox/q111402.bps}}}
\centerline{\epsfxsize=2.8in{\epsfbox{figures/qcurves/bbox/q111.bps}}}
\caption{The neutral stability curve for the basic state and the numerical saddle-nodes
for squares (solid squares) are shown as a functions of the wavenumber $q$. Open squares
denote cases where the single square loses stability. For the parameters
$ b = c = 1.0, e = 4.0, \beta_1 = 0.0, \beta_2 = 2.3$ (a), the oscillons
exist for values of the control parameter below the saddle-node for squares.  For the
parameters $ b = c = 1.0, e = .75, \beta_1 = \beta_2 = 2.0$ (b), the oscillons
no longer exist before the saddle-node for squares is reached by decreasing $R$.}
\label{fig:qcurves}
\end{figure}

To investigate whether the general locking mechanism is sufficient to explain the 
experiments or whether aspects have to be taken into account that are specific to the
granular system, we compare our findings with the experimental results in more detail.
The experimental phase diagram in Fig.~\ref{fig:phasediag} \cite{UmMe96} shows that oscillons 
exist only below the 
patterns for a mid-range of frequencies.  Increasing the driving amplitude $\Gamma$,
patterns appear at the solid diamonds.  For decreasing $\Gamma$, extended patterns disappear at
the solid circles to become either the flat layer or oscillons which exist all the way to the solid
squares.
Within the minimal framework presented here,
oscillons should exist where squares are strongly subcritical.  However, in the low 
frequency range where squares exhibit strong hysteresis, oscillons are not seen
experimentally. We propose the following explanation.  In the experiments, the
driving amplitude is increased into the pattern regime which results in a pattern 
with wavenumber $q \approx q_c$.  Then the amplitude is decreased and oscillons emerge
depending on where the pattern with this wave number goes unstable.  Fig.~\ref{fig:qcurves} 
shows the neutral stability curve for the basic state $\psi \equiv 0$
and the numerically determined range over which square patterns (restricted to one wavelength) are stable 
for
two sets of parameter values.  Decreasing $R$ below the value marked by solid squares, the square pattern
goes to zero in a saddle node bifurcation.  The open squares indicate the value of $R$ where the square pattern
seems to undergo an instability involving the wavelength of the pattern.
Note that we do not associate a wavenumber $q$ with the oscillon and the shaded region refers only
to the $R$-values.
As the control parameter $R$ is decreased, squares remain in
existence until the solid saddle-node points, although they may lose stability
earlier.  In Fig.~\ref{fig:qcurves}a the oscillon range 
extends below the saddle-node for squares, so that oscillons are expected as
$R$ is decreased below the existence of squares.
In contrast, Fig.~\ref{fig:qcurves}b shows a situation in which the oscillon range 
is above these points near $q_c = 1$,
and we do not expect to obtain oscillons by decreasing $R$ further below the saddle-node.

\begin{figure}[ht]
\centerline{\epsfxsize=2.8in{\epsfbox{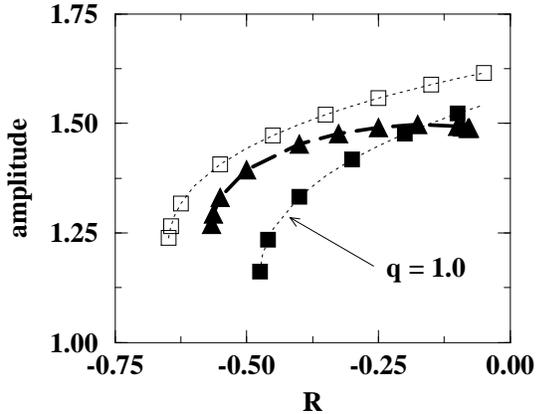}}}
\caption{ Bifurcation curves obtained for a single oscillon (solid triangles) and squares with
$q = 1.0$ (solid squares) and $q = 0.875$ (open squares).
Parameters are $b = c = 1.0, e = 4.0, \beta_1 = 0.0, \beta_2 = 2.3$ (cf. Fig.~\protect\ref{fig:qcurves}a). }
\label{fig:bifcurves}
\end{figure}

\begin{figure}[ht]
\centerline{\epsfxsize=1.in{\epsfbox{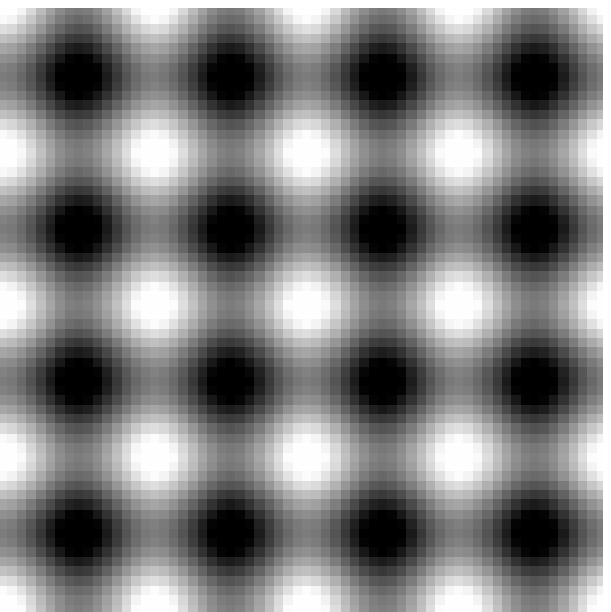}}\hspace{.1in}
       \epsfxsize=1.in{\epsfbox{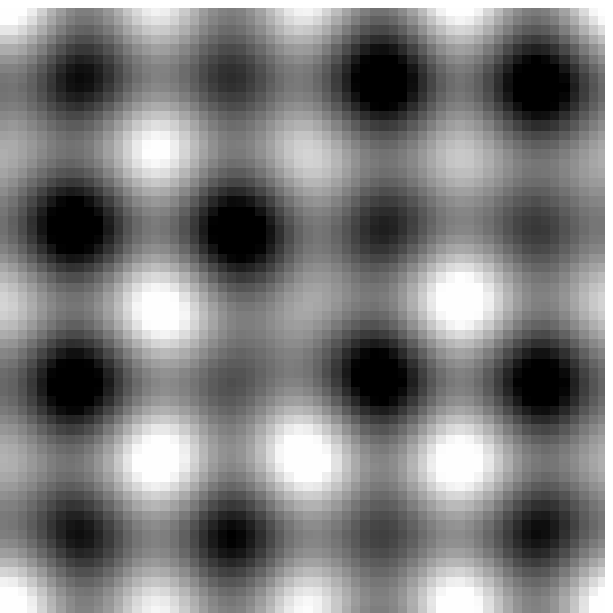}}\hspace{.1in}
       \epsfxsize=1.in{\epsfbox{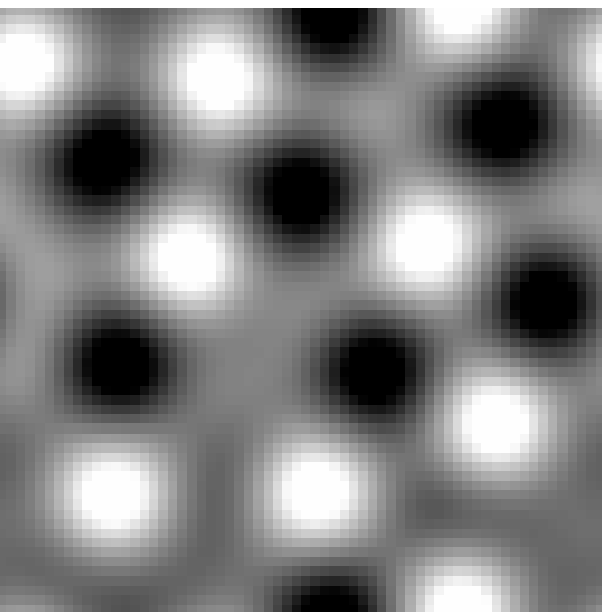}}}
       \vspace{.3in}
       \centerline{\epsfxsize=1.in{\epsfbox{figures/quench/a.504onet0.ps}}\hspace{.1in}
       \epsfxsize=1.in{\epsfbox{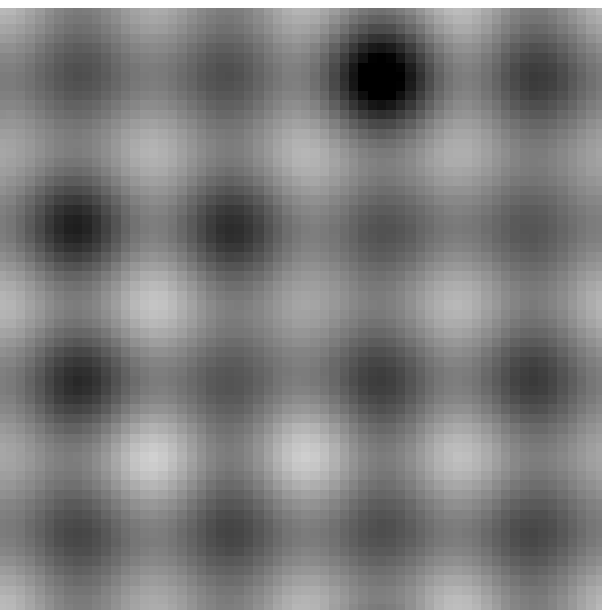}}\hspace{.1in}
       \epsfxsize=1.in{\epsfbox{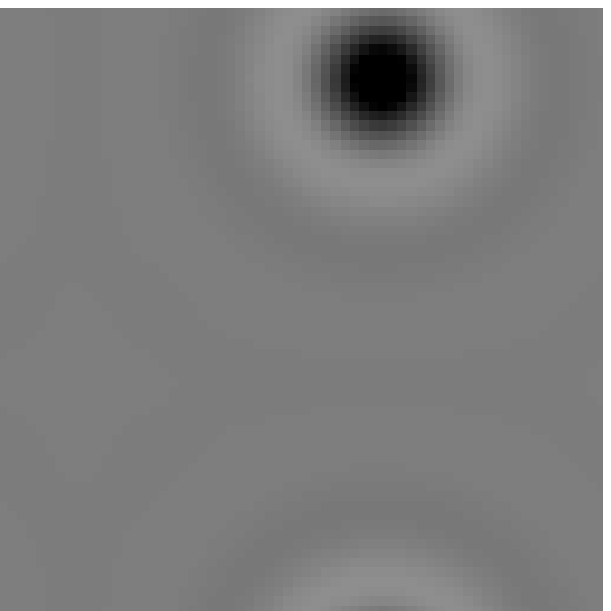}}}
\caption{(a) An initially perturbed square pattern with $R = -0.45$, then $R$ is decreased to $-0.48$
	 (b) where parts of the pattern decay ($t = 15$), and (c) several oscillons persist ($t = 1000$).  
	 (d) Again,
	 the perturbed square pattern with $R = -0.45$, (e) followed by rapid decay of the
	 pattern for $R = -0.504$ ($t = 10$), (c) where only one oscillon survives ($t = 2000$).
         All other parameters as in Fig~\protect \ref{fig:qcurves}a. }
\label{fig:quench}
\end{figure}

Fig.~\ref{fig:bifcurves} shows the bifurcation diagram obtained from numerical simulations of (\ref{e:sh})
for the same parameters as in Fig.~\ref{fig:qcurves}a.  
We note that stripes are supercritical in this case.
Here we see that a single oscillon exists for a range of $R$ values below the saddle-node for the
square pattern with wavenumber $q = 1$.
Starting with a slightly perturbed, extended square pattern
and then decreasing
the control parameter below the saddle-node, we find that the pattern breaks down and some
oscillons persist.  Figs.~\ref{fig:quench}a-c show a case in which several oscillons survive, while
Figs.~\ref{fig:quench}d-f show a case in which only one oscillon remains.
These results are consistent with the observation 
of oscillons existing below the periodic patterns in the granular experiments.  

In light of the experiments for lower frequencies, the oscillons may still exist, but
their range of existence may be `embedded' in the range of square existence as suggested
by Fig.~\ref{fig:qcurves}b.  In that case the oscillons would not be accessible by changing the
amplitude of forcing.  We suggest that oscillons may be accessible in this regime by
starting with an oscillon in the mid-frequency range and then decreasing the frequency. 

\section{Interaction of Oscillons}

\begin{figure}[ht]
\centerline{\epsfxsize=1.in{\epsfbox{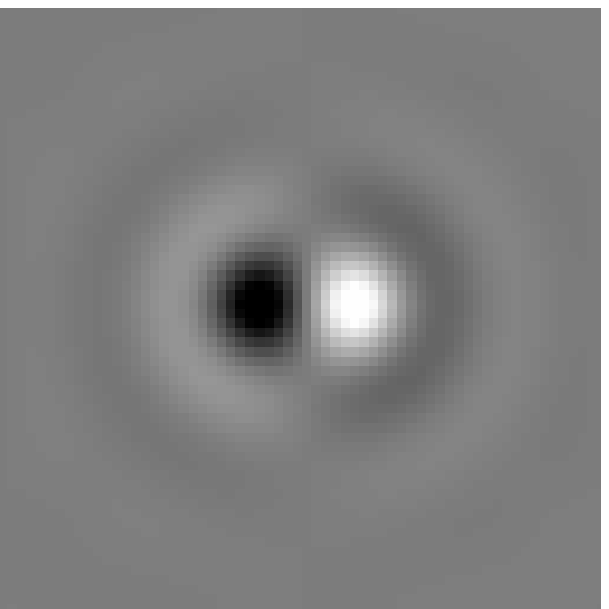}}\hspace{.1in}
       \epsfxsize=1.in{\epsfbox{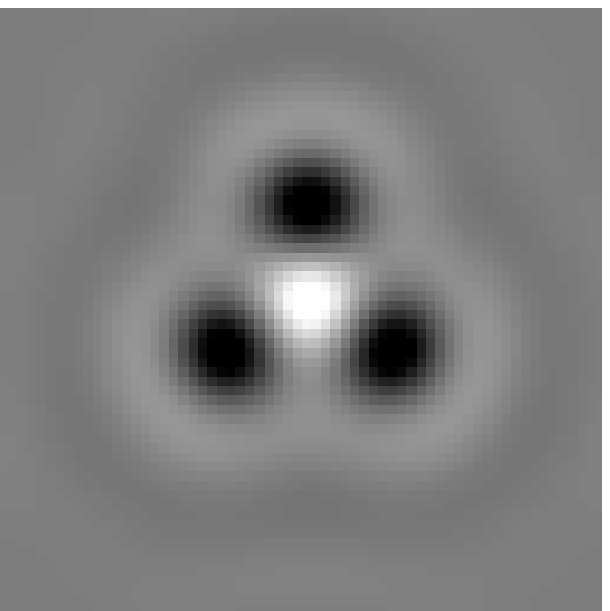}}\hspace{.1in}
       \epsfxsize=1.in{\epsfbox{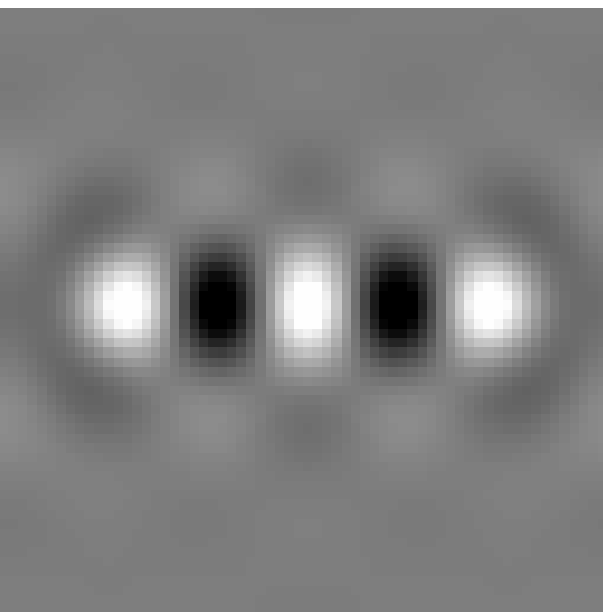}}}
\caption{Bound states of opposite polarity:  (a) dipole, (b) tetramer, and (c) chain of 5 alternating
         oscillons for the parameters as in Fig.~\protect \ref{fig:oscillon}.  }
\label{fig:bound}
\end{figure}

Since the oscillons have oscillatory tails that decay exponentially outward, 
one expects them to interact and form bound states.
Fig.~\ref{fig:bound} shows several of these bound states found in 
the numerical simulations which are similar to states observed in the experimental system.
These include dipoles, tetramers, and chains of alternating polarity.

\begin{figure}[ht]
\centerline{\epsfxsize=2.in{\epsfbox{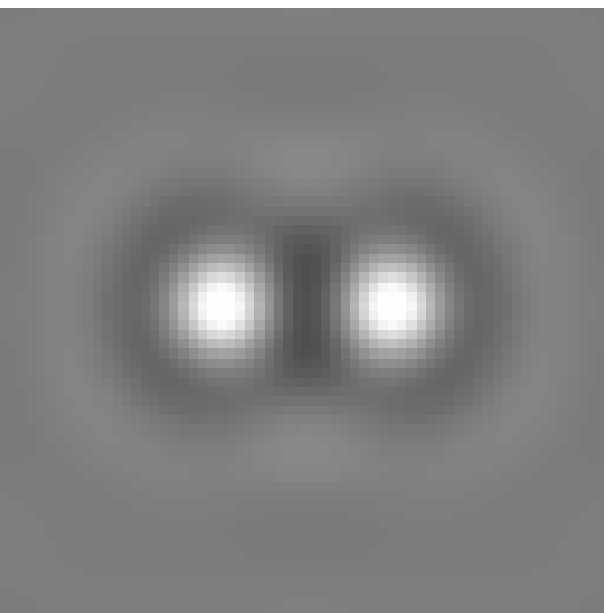}}}
\centerline{\epsfxsize=2.8in{\epsfbox{figures/pairs/bbox/crossection.bps}}}
\caption{(a) Two bound oscillons of like polarity and (b) their cross-section (solid line)
along with the cross-section for a chain of alternating polarity (dashed line).
For the pair $R = -0.470$ while for the chain $R = -0.469$. All other parameters as in Fig.~\protect
\ref{fig:oscillon}.}
\label{fig:pp}
\end{figure}

In addition to seeing oscillons of opposite polarity binding, we also find bound states of 
like polarity, which have not been seen in the experiments.  The separation
distance between like pairs is greater than for opposite pairs.  
As can be seen in the cross-section of a like pair, Fig.~\ref{fig:pp}, 
the spatial oscillations 
between the pair lock the two oscillons together, but do not create a third negative oscillon
between them as can be seen by comparison with the $+-+$ chain (dashed line).  However, increasing the 
parameter $R$ causes the oscillations to decay slower in space.
Eventually a value of $R$ is reached where the minimum  
exceeds a threshold beyond which the amplitude grows to form an oscillon
of opposity polarity and becomes a $+-+$ chain.

\begin{figure}[ht]
\centerline{\epsfxsize=2.8in{\epsfbox{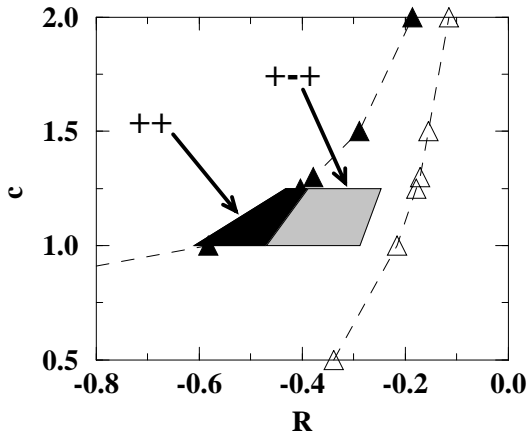}}}
\caption{ Existence range for a pair of like oscillons is shown in black.  The
left side of the grey region marks the transition to a chain of three oscillons of
alternating polarity (cf. Fig. \protect \ref{fig:pp}b).  The $+-+$ chain persists up to the 
right end of the grey area.
Parameters are as in Fig.~\protect\ref{fig:range}.}
\label{fig:pprange}
\end{figure}

Fig.~\ref{fig:pprange} shows the existence range of $++$ structures for two values
of $c$ (dark region).
The border between the dark and grey regions marks the transition to a $+-+$
structure.  To the right of the grey region the $+-+$ chain evolves to a periodic pattern.
This result shows that
the ratio of the like-pair range to the single-oscillon range is
greatly reduced by slightly increasing $c$.  Increasing $c$ further makes it difficult to find a
stable $++$ structure.  We expect that a bound pair of like polarity may be similarly difficult
to find in experiments because they exist only over a small range.  Conversely their range of
existence may be larger in regions of stronger subcriticality.

\begin{figure}[ht]
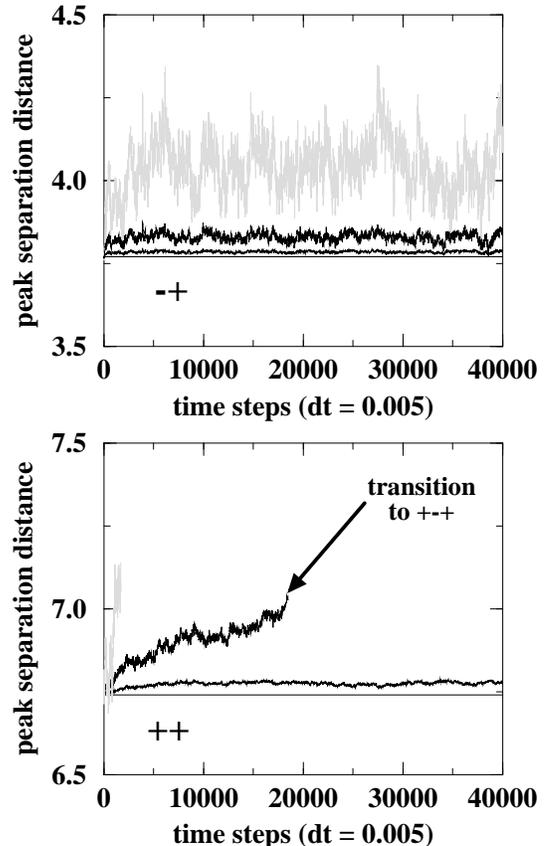

\centerline{\epsfxsize=2.8in{\epsfbox{figures/noise/bbox/npnoise.bps}}}
\centerline{\epsfxsize=2.8in{\epsfbox{figures/noise/bbox/ppnoise.bps}}}
\caption{ (a) The peak-peak distance for a pair of oscillons of opposite polarity is plotted for 
the unperturbed
pair (straight line) and for three added noise levels, $0.075$ (bottom), $0.15$ (middle), $0.3$
(top, grey).  (b) The same is plotted for a like pair.
The parameters are as in Fig.~\protect\ref{fig:oscillon}.}
\label{fig:noise}
\end{figure}

Since the peak-peak distance for two like pulses is greater than that for two pulses
of opposing amplitude, we expect the binding between the $++$ state to be much weaker   
than the binding of the $-+$ state.
In order to study the `strength' of the binding, we add bounded noise to the solution at each
time step and measure the peak-peak distance.  The results for several noise strengths
are shown in Fig.~\ref{fig:noise}.  Fig.~\ref{fig:noise}a shows that the opposing
pair remains bound for all three noise levels.  In contrast, Fig.~\ref{fig:noise}b shows that 
the like pair is more susceptible to the same values of noise.  In fact, for the
mid-level noise, the like pair creates a third negative oscillon in the center and for
the highest-level noise, the pair quickly loses
stability and by 40000 time steps has become 
two perpendicular chains of alternating polarity.  The effect
of noise in the experimental system combined with the small existence range for oscillons
of like polarity may contribute to them not being observed in the experiments.

\begin{figure}[ht]
\centerline{\epsfxsize=2.in{\epsfbox{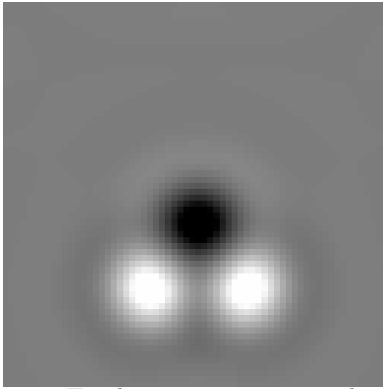}}
}
\caption{Triad propagating downward for $R = -1.5, b = c = 1.0, e = \beta_2 = 4.0, \beta_1 = 0.0$.}
\label{fig:triad}
\end{figure}

The interaction of like pairs can also result
in forming triad structures of two like oscillons bound to a third opposite oscillon in a 
triangular pattern as shown in Fig.~\ref{fig:triad}.  The triad breaks the spatial reflection symmetry 
and in this non-potential case the triad travels in the direction of the two
leading oscillons.

\section{Conclusion}

We have presented the results of investigating a Swift-Hohenberg model for the existence
of two-dimensional localized oscillons.  Oscillon-type structures are a general feature 
to be expected in systems where the transition to squares is sufficiently subcritical so as
to allow localization caused by the interaction of the large-scale envelope with the small-scale 
underlying periodic pattern.
Hence the observance of oscillons
should not be restricted to granular materials.  For example, we would also expect them to
be observed in Faraday experiments of shear-thinning fluids for which the viscosity decreases
with the flow, which should make the transition subcritical.  Since the subharmonic 
response results in the reflection symmetry $\psi \rightarrow -\psi$, one obtains equal
oscillons of opposite polarity.  For transcritical hexagons similar
localized structures are also obtained, but since there is no reflection symmetry, one does 
not obtain two different, equivalent oscillon-structures \cite{ArGo90}.

A comparison with the granular experiments will show whether this general mechanism
is sufficient to capture the observed oscillon.  Experimental tests to determine the validity of
this minimal framework involve the range of existence of oscillons and of like-pairs 
in the low frequency
range.  We expect that oscillons could be observed experimentally in the low frequency regime
by changing the experimental protocol.
We also note that bound states of like-polarity are shown numerically to be not as
robust as those of opposite polarity.  Their range of existence may be larger in
regions of greater subcriticality, e.g. lower frequencies, and their observance more likely.

The authors gratefully acknowledge discussions with I. Aranson, C. Bizon, B. Kath, 
M. Shattuck, L. Tsimring, and P. Umbanhowar.  
This research has been supported by DOE under grant DE-FG02-92ER14303.

\bibliography{/users/cath/.bibfiles/journal}

\end{document}